# Signal prediction by anticipatory relaxation dynamics

Henning U. Voss

*Weill Cornell Medical College, Citigroup Biomedical Imaging Center*
*516 East 72nd Street, New York, NY10021, USA*
*e-mail address: hev2006@med.cornell.edu*

Real-time prediction of signals is a task often encountered in control problems as well as by living systems. Here a model-free prediction approach based on the coupling of a linear relaxation-delay system to a smooth, stationary signal is described. The resulting anticipatory relaxation dynamics (ARD) is a frequency-dependent predictor of future signal values. ARD not only approximately predicts signals on average but can anticipate the occurrence of signal peaks, too. This can be explained by recognizing ARD as an input/output system with negative group delay. It is completely characterized, including its prediction horizon, by its analytically given frequency response function.

Real-time prediction of signals plays an important role in applied physics, for example electronic circuits used for signal processing and control, as well as in living systems, for example in cognitive motor control. It requires an instant computation of future signal values. Whereas most prediction mechanisms conceived so far require a model of the signal to be predicted, or, if such a model does not exist, an estimate of a model derived from the signal itself [1], physics recently has provided two apparently distinct approaches that enable real-time prediction of signals in an entirely different way: (i) Negative group delay: Negative group delay is observed as an apparent advancement of signal peaks in media with anomalous dispersion, some electronic filters, and metamaterials [2-6]. This counterintuitive behavior can be understood by the transfer or frequency response functions of these systems. (ii) Synchronization between dynamical systems: Some systems can be predicted in real time by coupling identical system copies to them, if the coupling term contains a time-delayed state of the coupled system. This anticipatory coupling then causes the systems to synchronize [7,8] in a way that the driven system leads the driving system in time, which is called anticipatory synchronization [9,10]. Although somewhat counterintuitive, too, the causal mechanism behind anticipatory synchronization is well understood [10,11].

Here it is shown that smooth, stationary, but not necessarily deterministic, signals can be predicted in real time by anticipatory coupling of a linear relaxation system. In this approach, anticipatory relaxation dynamics (ARD), modeling the signal to be predicted is not required. It is demonstrated that ARD predicts signals not only on average but that it also can predict the occurrence of signal peaks. This phenomenon is then explained by the concept of negative group delay. These findings considerably expand the application field of anticipatory coupling, and the simplicity of this mechanism might enable applications for which no model of the dynamics is known, including smooth random signals.

The article is organized as follows: First, prediction with ARD is shown in a numerical example. Then, a heuristic, synchronization based, explanation for prediction is provided. Although useful for the understanding of the anticipatory coupling term in ARD, it does not completely explain the observations and will be replaced by an accurate explanation based on negative group delay in linear systems. A physiological example and a discussion conclude the paper.

The ARD system is defined as follows: Let $x(t) \in \mathbb{R}$ ($t \in \mathbb{R}$) be a smooth, stationary signal with zero mean. For the sake of simplicity the signal is assumed to be sampled quasi-continuously in time, i.e., with time increments comparable to those one would use for numerical solutions of ordinary differential equation-based models of the signal. ARD is generated by the driven relaxation-delay system

$$\dot{y}(t) = -\alpha y(t) + K \cdot [x(t) - y(t-\tau)], \quad (1)$$

with parameters $\alpha, K \in \mathbb{R}$. The anticipatory coupling term $[x(t) - y(t-\tau)]$ consists of the difference between the input signal $x$ at time $t$ and the ARD signal $y$ at time $t-\tau$ with $\tau > 0$. Signal prediction is performed by numerically solving Eq. (1). The solution $y(t)$ is a predictor for the signal $x(t)$ as will be specified in the following numerical example.

*Example I:* In order to simulate a smooth random signal $x$, 1000 s of uniformly distributed random numbers are generated with a sampling rate of 1000 Hz. This signal is filtered with a (causal) Butterworth filter with a cutoff frequency of 0.1 Hz, resulting in a smooth, or band-limited, signal. Equation (1) is numerically integrated with $\tau = 2$ s, $K = 5.4$ Hz, and $\alpha = 5.3$ Hz. The signal $x$ and the predicted, or anticipated, signal $y$ are shown in Fig. 1a. The anticipation time, estimated by the negative argument of the maximum of the cross-correlation function $C_{xy}(\tau)$ between $x$ and $y$, is $\tau_C = 0.91$ s (Fig. 1b), and $C_{xy}(-0.91) = 0.99$. It turns out that 91% of signal peaks are predicted accurately (Fig. 1e) with a mean peak anticipation time of $\tau_P = 0.73$ s. Computational details are given at the end of the second example.

In order to understand how ARD predicts a signal, first a heuristic and then an accurate explanation is provided.

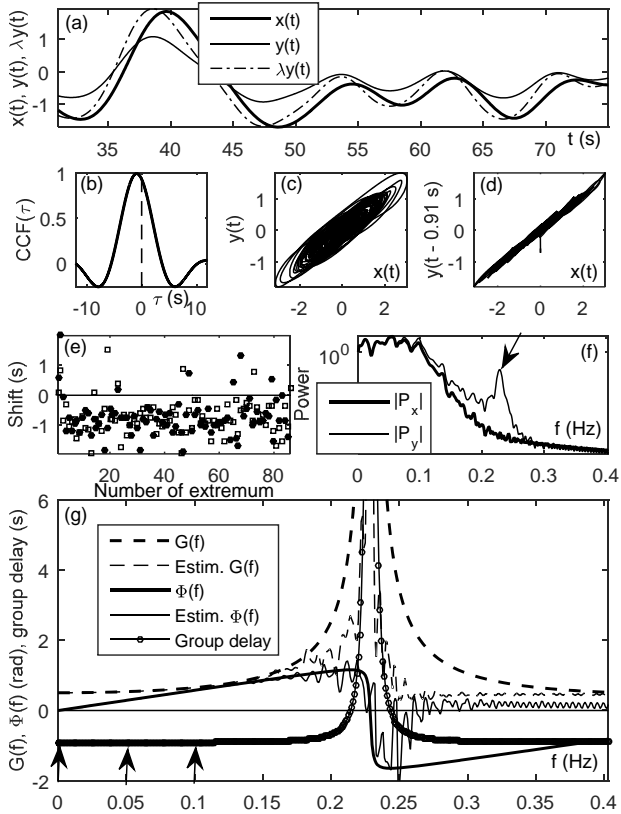

FIG. 1: Anticipatory relaxation dynamics (ARD) for a smoothed random signal. a) Section of the signal to be predicted, $x$, and the predicted ARD signal, $y$. The rescaled signal $\lambda y$, with $\lambda$ a scaling factor, is shown, too. b) Cross-correlation function between $x$ and $y$. The dashed line marks $\tau = 0$. c) Phase portrait of $x(t)$ vs. $y(t)$. d) Phase portrait of $x(t)$ vs. $y(t - \tau_C)$, demonstrating attraction to the anticipation manifold, defined as $x(t) = \lambda y(t - \tau_C)$, after a short transient. e) Shift of predicted signal peaks with respect to actual signal peaks. Open boxes denote maxima and bullets denote minima. Negative shifts correspond to signal anticipation, positive shifts to signal lag. f) Logarithmically scaled normalized power spectrum of $x$ (bold line) and $y$ (line). The arrow points at increased out-of-band power of $y$. g) Analytic frequency response function gain $G$ and phase $\Phi$, from Eqs. (5) and (6), respectively, and their estimates from data, as well as the analytic group delay $\delta$. Estimates are very accurate for low frequencies and less accurate for $f \gg 0.1$ Hz due to the lack of out-of-band signal power. The arrows point to the first frequency band with negative group delay. The maximum gain (outside of plot area) is 59.

(i) Heuristic explanation: Define *formally* a non-autonomous dynamical system with state $\tilde{x}$ that copies the signal $x$ by

$$\dot{\tilde{x}}(t) = -\alpha \tilde{x}(t) + f(t), \qquad (2)$$

with $f(t) = \alpha \tilde{x}(t) + \dot{\tilde{x}}(t)$. Replacing $x$ in Eq. (1) with $\tilde{x}$, the coupling term $K \cdot [\tilde{x}(t) - y(t - \tau)]$ ($\tau > 0$) is known as anticipatory coupling [9], for the following reason: The transverse system $\Delta(t) = \tilde{x}(t) - y(t - \tau)$ obeys the dynamics

$$\dot{\Delta}(t) = -\alpha \Delta(t) - K\Delta(t - \tau) + f(t), \qquad (3)$$

a non-autonomous linear delay-differential equation. First, consider the trivial but instructive case of an autonomous system with $f(t) = 0$. Independent of the parameters $\alpha$ and $K$, it has a trivial fixed point $\Delta(t) = 0$. But in this fixed point, $\tilde{x}(t + \tau) - y(t) = 0$, or $y(t) = \tilde{x}(t + \tau)$. Therefore, the ARD system (1) anticipates, or predicts, the signal $\tilde{x}$ at a time $t + \tau$. The stability conditions of this fixed point are well known ([12], reviewed in [13,14]), and if stable, the fixed point, or equivalently, anticipation manifold $\tilde{x}(t) = y(t - \tau)$, is globally attractive. Now consider the more realistic case, which is Eq. (3) including the disturbance $f(t)$. The disturbance works against the attraction of the transverse system to its origin at $\Delta(t) = 0$, and thus prevents complete settlement on the anticipation manifold. Therefore, a better understanding of the mechanism of prediction by ARD is required. It is provided in the following.

(ii) Explanation by negative group delays: Since the dynamics of $y$ is linear in $x$, it can be analyzed by the frequency response function of the ARD system. With $f$ being frequency, $\omega = 2\pi f$, $x(t) = \int X(\omega) e^{i\omega t} d\omega$, and $y(t) = \int Y(\omega) e^{i\omega t} d\omega$, the input/output relationship between $x$ and $y$ is given by $Y(\omega) = H(\omega) X(\omega)$, in which $H(\omega)$ is the frequency response function of the ARD system (1). It is

$$H(\omega) = G(\omega) e^{i\Phi(\omega)} = \frac{K}{\alpha + i\omega + K e^{-i\omega\tau}} \qquad (4)$$

$$= \frac{K}{\beta}[\alpha + K\cos(\omega\tau)] + i\frac{K}{\beta}[K\sin(\omega\tau) - \omega],$$

with $\beta = [\alpha + K\cos(\omega\tau)]^2 + [\omega - K\sin(\omega\tau)]^2$. Its gain and phase are, respectively,

$$G(\omega) = K/\sqrt{\beta}, \qquad (5)$$

$$\Phi(\omega) = \arg\{\alpha + K\cos(\omega\tau) + i[K\sin(\omega\tau) - \omega]\}. \qquad (6)$$

In Eq. (4), the term $e^{-i\omega\tau}$ in the denominator causes an anticipatory time shift [15] and the term $\alpha + i\omega$ the dissipative dynamics of the ARD system. The actual anticipation time induced by the ARD system now is frequency dependent and given by

$$\tau_\omega(\omega) = \frac{d\Phi(\omega)}{d\omega} \tag{7}$$

$$= -[K\cos(\omega\tau) - K^2\tau - \alpha(K\tau\cos(\omega\tau) - 1) + K\tau\omega\sin(\omega\tau)]/\beta .$$

In case of anticipatory dynamics, $\tau_\omega(\omega) > 0$, taking account for the convention of positive delay times in Eq. (1). The expression $\delta(\omega) = - d\Phi(\omega)/d\omega$ is usually called the group delay of a signal, and negative group delay in general means a group advance [3,4], or anticipation in the case of ARD.

Proceeding with the numerical example, it is now possible to estimate the mean anticipation time, $\tau_a$, from data and to compare it with the mean of the analytic result (7). Numerical estimates of the frequency response phase spectrum and gain, and their comparison with the analytic expressions (6) and (5), respectively, are shown in Fig. 1g. Within the frequency band of the low pass filter, [0, 0.1] Hz, the phase spectrum is approximately linear (arrows), allowing for a linear regression fit of $\tau_a$ over this frequency band. Its estimate is $\tau_a = 0.91$ s, equaling $\tau_C$ obtained before. It also matches the analytic anticipation times (7), which are approximately constant within this frequency band (Fig. 1 g). Simulations with larger data sets (not shown) suggest that for the chosen parameters asymptotically $\tau_C \approx \tau_a \approx \tau_\omega$. As $\tau_\omega$ can be computed via Eq. (7), it is completely determined by $\alpha$, $K$, and $\tau$.

Distortion of signals is an expected phenomenon in systems with negative group delay [4,16]. Distortions and the observation that $\tau_C < \tau$ requires a phenomenological definition of the anticipation manifold for ARD. In this example, defining the anticipation manifold as $x(t) = \lambda y(t - \tau_C)$ provides a reasonable approximation (Figs. 1c,d). The scaling factor $\lambda$ can be obtained by matching the variances of $x$ and $y$ (Fig. 1a). Finally, Fig. 1f shows that for frequencies between about 0.15 and 0.3 Hz the power spectrum of $y$ increases relative to the respective $x$ power spectrum components (arrow), which results from the high gain of the frequency response function for these out-of-band frequencies (Fig. 1g). For other parameter choices, these high gains can cause for example oscillatory instabilities [16].

Adding white noise to the signal and keeping the ARD parameters unchanged shows that ARD seems to be quite insensitive to moderate amounts of signal noise. For example, additive white noise with 50% of signal standard deviation to $x$ provides again $\tau_C = 0.91$, with $Cxy$(-0.91) = 0.88.

It is evident that the ARD frequency response function resembles those of some electronic circuits and metamaterials with negative group delay (Fig. 2 of Ref. [3] and Fig. 1 of Ref. [6], respectively): a frequency band with negative group delay followed by a band with positive group delay. Therefore, some frequency components of $x$ are advanced, some are delayed by $y$. The consequences of this will be elaborated in the next example.

*Example II:* In order to demonstrate applicability to naturally variable signals with more complex power spectra, $T = 60$ s of a cardiac pulse signal obtained with photoplethysmography from the finger of a healthy volunteer (part of the public data set [17]) is predicted with ARD. The signal, sampled at a rate of 100 Hz, is first interpolated to 1 kHz sampling rate. Equation (1) is integrated with $\tau = 150$ ms, $K = 105$ Hz, and $\alpha = 1.5K$. Figure 2a shows a section of the result. The anticipation time is $\tau_C = 31$ ms, with $Cxy$(-31 ms) = 0.87. A linear regression fit to the phase spectrum over the frequency band that includes the main peak of the power spectrum of $x$ (Fig. 2g, arrows) gives $\tau_a = 48$ ms. However, for the second and third largest peaks of the power spectrum, which reflect the dynamics of the minor minima in the time series (arrow in Fig. 2a), the group delay becomes positive, so these frequency components are lagging. Further, the corresponding power spectrum components of $y$ increase (arrow in Fig. 2f), which is again caused by the high gain for these frequencies (Fig. 2g). Positive group delay and high gain are responsible for the observation that the minor minima are not anticipated but lagging, whereas all maxima and most principal minima are predicted after a short transient time (Fig. 2e). The mean peak anticipation time is $\tau_P = 30$ ms. From the phase portraits in Figs. 2c,d it is evident, too, that the minor minima are lying off the anticipation manifold.

(Computational details: The parameters $\tau$, $K$, and $\alpha$ were optimized with a graphical user interface showing their effect on the frequency response function and group delay. It is provided as Supplemental Material at www.mathworks.com/matlabcentral/fileexchange/52435-ard-frequencyresponse-gui--. This GUI can also be used to verify the insensitivity of the frequency response function against small parameter changes. Peak counting: Since the number of peaks in $y$ can exceed the number of peaks in $x$ due to an increase in signal power for high frequencies as described above, peaks in $y$ that exceeded a distance threshold to peaks in $x$ were removed, and then shifts of peaks in $y$ with respect to corresponding peaks in $x$ were assigned a negative value if the peak in $x$ was predicted by the peak in $y$. The signal was transformed to zero mean and unit standard deviation before integrating Eq. (1). Equation (1) was solved with a Runge-Kutta scheme of $4^{th}$ order. The initial conditions were $y(0) = x(0)$ and $y(t) = 0$ for $t \in$ [-$\tau$, 0). All computations were performed with MATLAB R2015a (The MathWorks, Inc., Natick, MA), including its cpsd function to estimate power and phase spectra.)

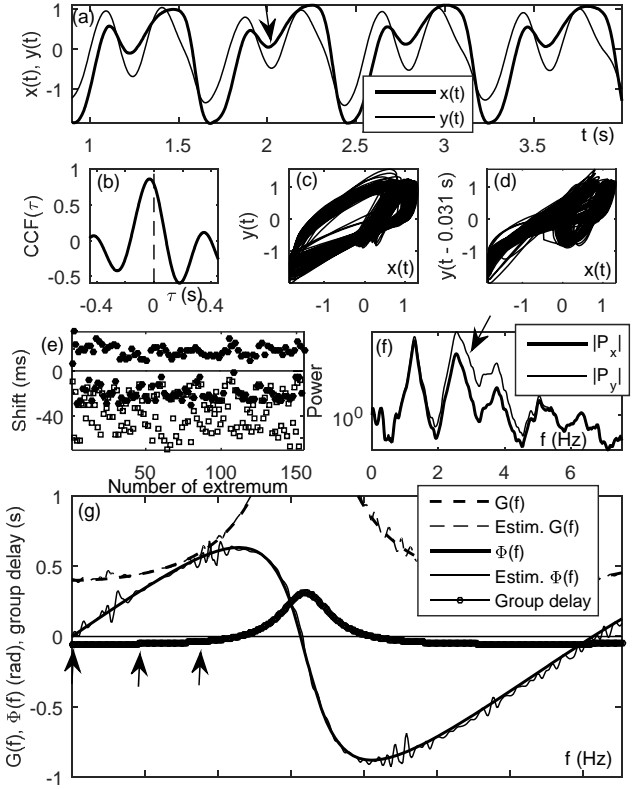

FIG. 2: Like Fig. 1 but for a physiological signal, a human cardiac pulse time series. a) Section of the signal and the predicted signal. The arrow points at a minor minimum. e) Almost all maxima and all major minima are anticipated, whereas minor minima are not anticipated (positive shifts). g) The arrows point to the first frequency band with negative group delay. The maximum gain is 2.0. Please refer to caption of Fig. 1 for further description of this figure.

A discussion concluders this paper: An important difference of ARD to anticipatory synchronization [9,10] is that in the latter, $\tau_C \approx \tau$ is to be expected [18], and has been observed even for stochastic forcing of the master system [19]. The inequality $\tau_C < \tau$ reflects that ARD cannot completely settle on the anticipation manifold, and that the group delay is frequency dependent. Therefore, although it provides a rationale for the specific coupling term used here, the heuristic, synchronization based derivation of ARD should not be overstressed. Rather, ARD is described entirely by its frequency response function, yielding negative group delay for certain frequencies only. In a similar sense, there might be more effective coupling terms than specified in Eq. (1) [20-25]. Cascading ARD systems might increase anticipation times at the cost of prediction accuracy or system stability [4,16].

In contrast to prediction schemes that estimate a model directly from past *signal* observations, such as nearest-neighbor-state methods [1], prediction by ARD relies on past values of the *predicted signal*. Therefore, there is no need for a memory of past signal values. However, the prediction horizon of ARD, $\tau_C$, seems to be more limited in comparison to model based approaches.

Anticipatory mechanisms for prediction and control are already used in physics and engineering [26-32], as well as in the neurosciences [33-44], but usually only work with signals generated by deterministic dynamical systems. Prediction with ARD will probably not be able to compete with model based or off-line schemes. But even if ARD prediction horizons are relatively short and predictions can be distorted, it still is conceivable that neuronal systems could have implemented ARD by utilizing synaptic or conduction delays, which are known to affect the dynamics of neuronal information processing [45-47]. In particular, it could be worthwhile to search for signatures of ARD in, for example, the distribution of peak prediction variability in cognitive motor control experiments [48]. Recently anticipatory dynamics in brain signaling have been observed and explained by time delays mediated by synaptic coupling properties [49]. Interestingly, the relationship $\tau_C < \tau$ recently has been observed in a manual tracking experiment of chaotic signals with external feedback delay [50], a task whose execution most likely does not involve an internal model of the signal system in untrained subjects. Finally, the frequency response function with delay, Eq. (4), although being known in the context of Langevin equations with delay [51], to the best of my knowledge has not been encountered in the design of systems with negative group delay yet and might inspire novel designs of control schemes or metamaterials.